# Hidden ordered compound-layer and its tailoring of the electronic/optical property in $Ge_2Sb_2Se_xTe_{5-x}$ alloys

*Chenxu Yu*[1], *Linggang Zhu*[1]*, *Hanyu Liu*[1], *Xianyao Huan*[1], *Naihua Miao*[1], *Jian Zhou*[1], *Zhimei Sun*[1]*

[1]School of Materials Science and Engineering, Beihang University, Beijing 100191, China.

*Corresponding authors: lgzhu7@buaa.edu.cn, zmsun@buaa.edu.cn

**Keywords:** $Ge_2Sb_2Se_xTe_{5-x}$ alloy, compound-layered structure, local chemical ordering, electronic property, optical property.

**Abstract**

$Ge_2Sb_2Se_xTe_{5-x}$ (0<x<5, GSST) alloys represent an emerging class of phase-change materials for integrated photonics. However, the microscopic origins underlying their superior performance compared to the parent compound $Ge_2Sb_2Te_5$ remain elusive. By using atomic simulations, this work elucidates that the thermal stability and low optical loss of GSST are fundamentally governed by the formation of an in-layer compound-like structure with $SeTe_2$ or $Se_2Te$ stoichiometry depending on the Se content, contrasting to the previously believed pure-element-layered model where Se and Te atoms occupy separate layers inside GSST. The newly identified compound-layered structures maintaining stability at temperature above 370 K, yield an enlarged bandgap, weakened antibonding character, and more importantly, a moderate refractive index as well as decreased extinction coefficient which align better with the experiment compared to the previously believed model. The present findings not only help bridge the long-standing theory-experiment gap regarding the optical properties of GSST by redefining its atomic structure, but also establish local chemical ordering as a critical materials design principle for high-performance photonics.

Chalcogenide phase-change materials (PCMs) have attracted extensive attention owing to their reversible amorphous-crystalline transition and the strong coupling between atomic structure and physical properties.[1-6] Among them, $Ge_2Sb_2Te_5$ (GST) and its derivatives are of particular interest because of their remarkable optical and electrical contrast between different phases, making them promising for non-volatile memory and integrated photonic applications.[7-9]

As a well-studied chalcogenide for optoelectronics, $Ge_2Sb_2Se_xTe_{5-x}$ ($0<x<5$, GSST) system, a doped derivative of the classic GST, by partially substituting Te with Se, has demonstrated superior properties over the parent GST. This composition modification significantly reduces optical loss,[10] enhances infrared transparency,[11] and improves thermal stability[12] and glass-forming ability[13]-thereby extending the durability and reliability of photonic devices.[14,15] The broadband infrared transparency of GSST further broadens its application potential in infrared optics and high-performance photonic computing.[16,17] Clearly, compositional optimization—specifically, the partial substitution of Te with chemically distinct Se—represents a primary consideration in explaining the enhanced properties of GSST compared to GST. Previous theoretical studies, which constructed GSST structural models by replacing one entire Te layer with Se, have demonstrated tunability in the optical properties relative to GST.[10,18,19] However, our preliminary calculations indicate that quantitatively such idealized GSST models, featuring perfectly segregated elemental layers (e.g., pure Te and pure Se layers, hereafter referred to as "element-layered structures"), yield optical constants that deviate significantly from experimental measurements. This discrepancy suggests that these previously believed models do not adequately capture the actual atomic-scale configurations of GSST.

Recently, it is found that for the doped and/or multicomponent semiconductors, in addition to the composition, local atomic arrangements can have significant influence on their electronic properties, such as carrier mobility, bandgap, mid-gap states and defect passivation.[20-23] Although the influence of local ordering on optical properties has received far less attention, we propose that it represents a new degree of freedom worthy of systematic exploration in optical material design. At present, overlooking possible local atomic ordering in GSST alloys likely results in an incomplete understanding of their fundamental physical behavior. Therefore, in this work, a thorough exploration of the stable atomic configurations of GSST is undertaken, aiming to bridge the gap between theoretical and experimental understanding of its optical properties. Furthermore, we will elucidate the role of local chemical ordering in regulating the optical property of the materials.

In this letter, four $Ge_2Sb_2Se_xTe_{5-x}$ systems with x=1, 2, 3, 4, are studied here, including $Ge_2Sb_2Se_1Te_4$, $Ge_2Sb_2Se_2Te_3$, $Ge_2Sb_2Se_3Te_2$ and $Ge_2Sb_2Se_4Te_1$, short for GSS1T4, GSS2T3, GSS3T2 and GSS4T1, respectively. Their stable atomic configurations are sampled using density functional theory informed Monte Carlo (MC) simulation, initiated from a quasi-random distribution of Se and Te atoms generated by using SQS method, detailed in the Supplementary Material Note-1. As the MC simulation proceeds, it is noticed that Te atoms tend to segregate onto the two atomic layers on either side of the vdW gap in the GSST structure, forming the Te-Te double layers. Thus to increase the efficiency of the simulation, the double layers near the vdW gap are fixed with Te atoms, while Se and Te atoms can freely populate in other original Te layers of GST (see Supplementary Material Note-2). Figure 1a summarizes the energetics of the structures with Se/Te randomly distributed (random) and Se/Te occupying distinguished atomic layers separately (element-layered), as well as the stable atomic configurations obtained by the MC simulations (compound-layered). As for the element-layered structures, the atomic layers in GSS1T4 are arranged as “Te-Ge-Se-Sb-Te-vdW-Te-Sb-Te-Ge-...”, while in GSS2T3 they are arranged as “Te-Ge-Se-Sb-Te-vdW-Te-Sb-Se-Ge-....”. Clearly, as shown in Figure 1a, for the GSS1T4 and GSS2T3 systems which correspond to a relatively low concentration doping of Se in $Ge_2Sb_2Te_5$, the compound-layered structures with Se/Te forming ordered compound-like structures in specific layers exhibit the highest stability, followed by element-layered configurations and then random structures. To exclude the influence of parameter setting on the energy calculation, a series of calculations using different combinations of exchange-correlation functionals and vdW correction schemes are performed (see Supplementary Material Note-3 for details), all confirming the stability of the compound-layered structures.

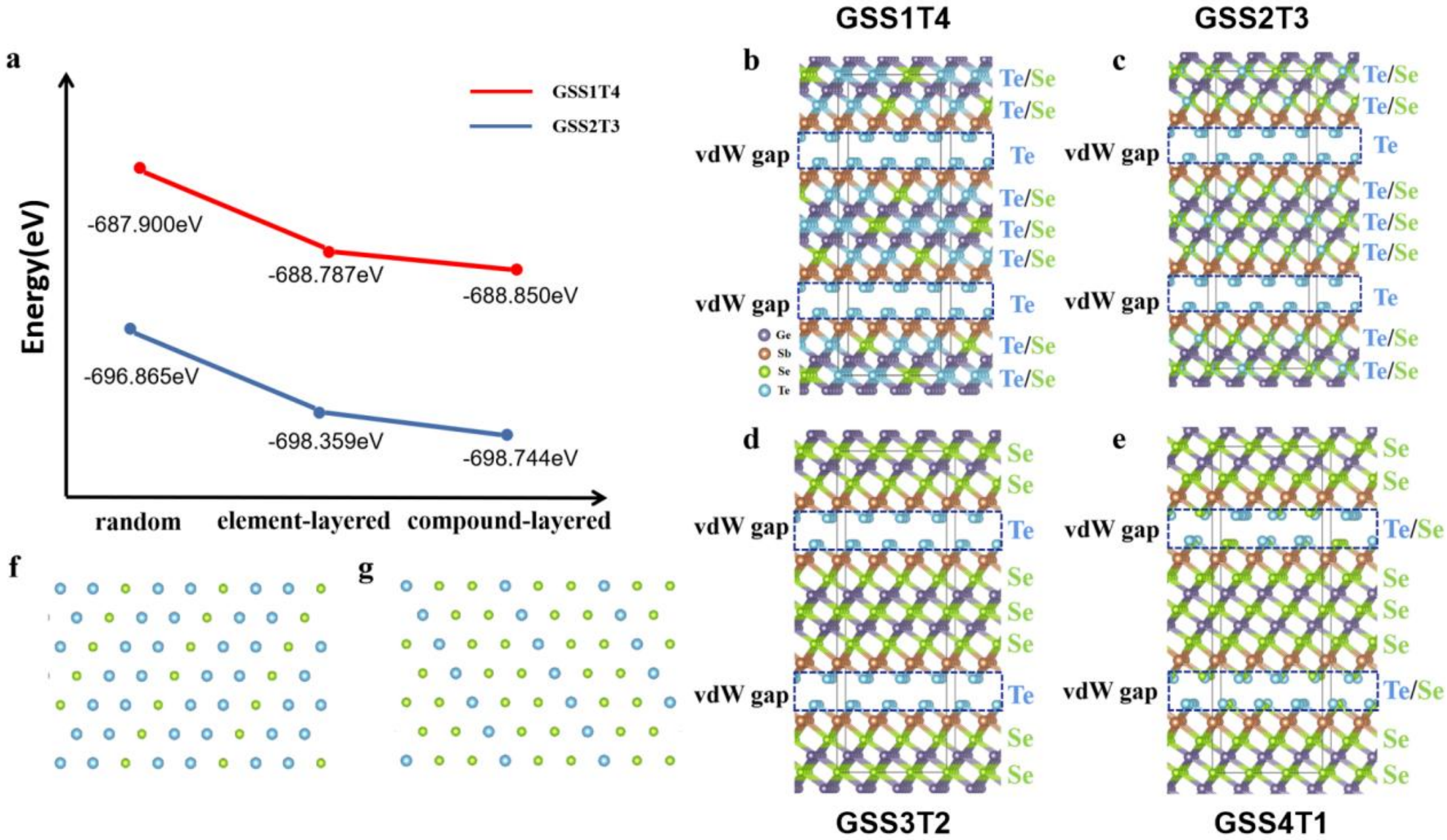

**Figure 1.** Structure energies and atomic configurations in $Ge_2Sb_2Se_xTe_{5-x}$. **a** Energy hierarchy of random, element-layered, and compound-layered configurations for $Ge_2Sb_2Se_1Te_4$ (GSS1T4) and $Ge_2Sb_2Se_2Te_3$ (GSS2T3). **b-e** The most stable configurations for GSS1T4, GSS2T3, GSS3T2 and GSS4T1, respectively. **f**, **g** In-layer compound-like Se/Te arrangement: $SeTe_2$ layer in GSS1T4 (**f**), $Se_2Te$ layer in GSS2T3 (**g**).

Diving into the detailed atomic configurations, as mentioned above, in contrast to the normally assumed element-layered structures where Se atoms occupy one entire layer between the Ge and Sb atomic planes, we find that the $Ge_2Sb_2Se_1Te_4$ and $Ge_2Sb_2Se_2Te_3$ systems exhibit a markedly different atomic arrangement (Figure1b-c): Se and Te atoms align with clear patterns, forming an ordered compound-like 2D structure with stoichiometry of $SeTe_2$ and $Se_2Te$, respectively, as depicted in Figure 1f-g. For $Ge_2Sb_2Se_3Te_2$, Se atoms fully occupy the original Te layers neighboring Ge/Sb layer, while Te forming a Te–Te double layer near the vdW gap, as shown in Figure 1d. When the content of Se further increases as in the case of $Ge_2Sb_2Se_4Te$, Se atoms still preferentially occupy Te layer neighboring Ge/Sb layer first, while excess Se atoms inevitably occupy sites within the Te-Te double layer near the vdW gap and mix randomly with Te, as denoted in Figure 1e. In the case of GSS3T2 and GSS4T1, the observed occupation preference of Se atom and its avoidance in the Te-Te double layers near the vdW gap are consistent with the reports in the literature.[10] As for the origin of the exclusion of Se in the Te-Te double layers, it can be attributed to the fact that Se incorporation into the Te-Te layer increases the interlayer distance, thereby weakening van der Waals interactions, as corroborated by the electron localization function (ELF) analysis which reveals a redistribution of electronic density and reduced interlayer cohesion upon Se incorporation (see Supplementary Material Figure S2).

The atomic configurations shown in Figure 1b-e can be thought as the ground-state structures for each composition of $Ge_2Sb_2Se_xTe_{5-x}$, in which Se/Te atoms are orderly distributed in specific layers rather than a random occupation. In practical environment with finite temperature, the entropy can contribute to the breaking of the chemical order, favoring a random arrangement of Se/Te.[24] To assess the thermodynamic stability of the relatively ordered structure found here, the temperature under which these structures will transform to a random state (with Se/Te randomly distributed) is calculated by taking into account the configurational entropy using the equations in the Supplementary Material Note-1. As shown in Table 1, the locally ordered compound-layers can still be preserved at temperatures of 370 K, indicating that these novel local structures can functionalize the GSST materials in practical devices. Moreover, it can be seen that the order-disorder transition temperature increases with increasing Se content, indicating a strengthening chemical bonding after Se doping in GST.

Based on the structural stability discussed above, the electronic properties of $Ge_2Sb_2Se_xTe_{5-x}$ with varying Se contents and different Se/Te configurations are further explored. As shown in Figure 2, the concentration dependence of the density of states and band structures is calculated based on the most-stable geometry for each composition. It shows that all $Ge_2Sb_2Se_xTe_{5-x}$ are narrow-bandgap semiconductors with a direct bandgap. Additionally, for pure $Ge_2Sb_2Te_5$, the calculated band gap is 0.25 eV at PBE level, in agreement with the literature.[25] And for $Ge_2Sb_2Se_5$, a band gap of 0.13 eV is obtained using the same functional setting, indicating that substituting Te with Se reduces the band gap. This trend still holds in quaternary $Ge_2Sb_2Se_xTe_{5-x}$, with the gradually decreasing gap from GSS1T4 to GSS4T1 as can be seen from Figure 2a.

**Table 1** Calculated enthalpy change (ΔH), configurational entropy change (ΔS), and order–disorder transition temperature (T) for $Ge_2Sb_2Se_xTe_{5-x}$ systems.

| Materials | **ΔH/eV** | **ΔS/eV·K$^{-1}$** | **T/K** |
|---|---|---|---|
| $Ge_2Sb_2Se_1Te_4$ | 1.21 | $3.23\times10^{-3}$ | 370 |
| $Ge_2Sb_2Se_2Te_3$ | 1.88 | $4.54\times10^{-3}$ | 414 |
| $Ge_2Sb_2Se_3Te_2$ | 2.08 | $4.54\times10^{-3}$ | 460 |
| $Ge_2Sb_2Se_4Te_1$ | 1.75 | $3.23\times10^{-3}$ | 541 |

To further investigate the impact of local structural ordering on the electronic properties, a comparative analysis among the compound-layered structure, element-layered structure, and random structure is conducted, focusing on the GSS1T4 and GSS2T3 systems, as displayed in Figure 2b and 2c. Compared with the random arrangement, the ordered structures (compound-layered structure and element-layered structure) exhibit an increased bandgap, and the compound-layered system show a slightly larger bandgap than the element-layered ones. The random configuration shows apparent carrier accumulation at the band edge including the valence band maximum (VBM) and conduction band minimum (CBM), diminishing the stability of the system. To clarify the specific contributions of Se/Te ordering to the electronic structure, the layer-resolved density of states contributed by Se/Te layer and the other layers (including Ge and Sb layer) are compared (Supplementary Material Figure S7). The results reveal that Se/Te layers provide the dominant contribution to the VBM, including the observed carrier accumulation at the band edges, underscoring the crucial role of local atomic geometry in shaping the electronic properties.

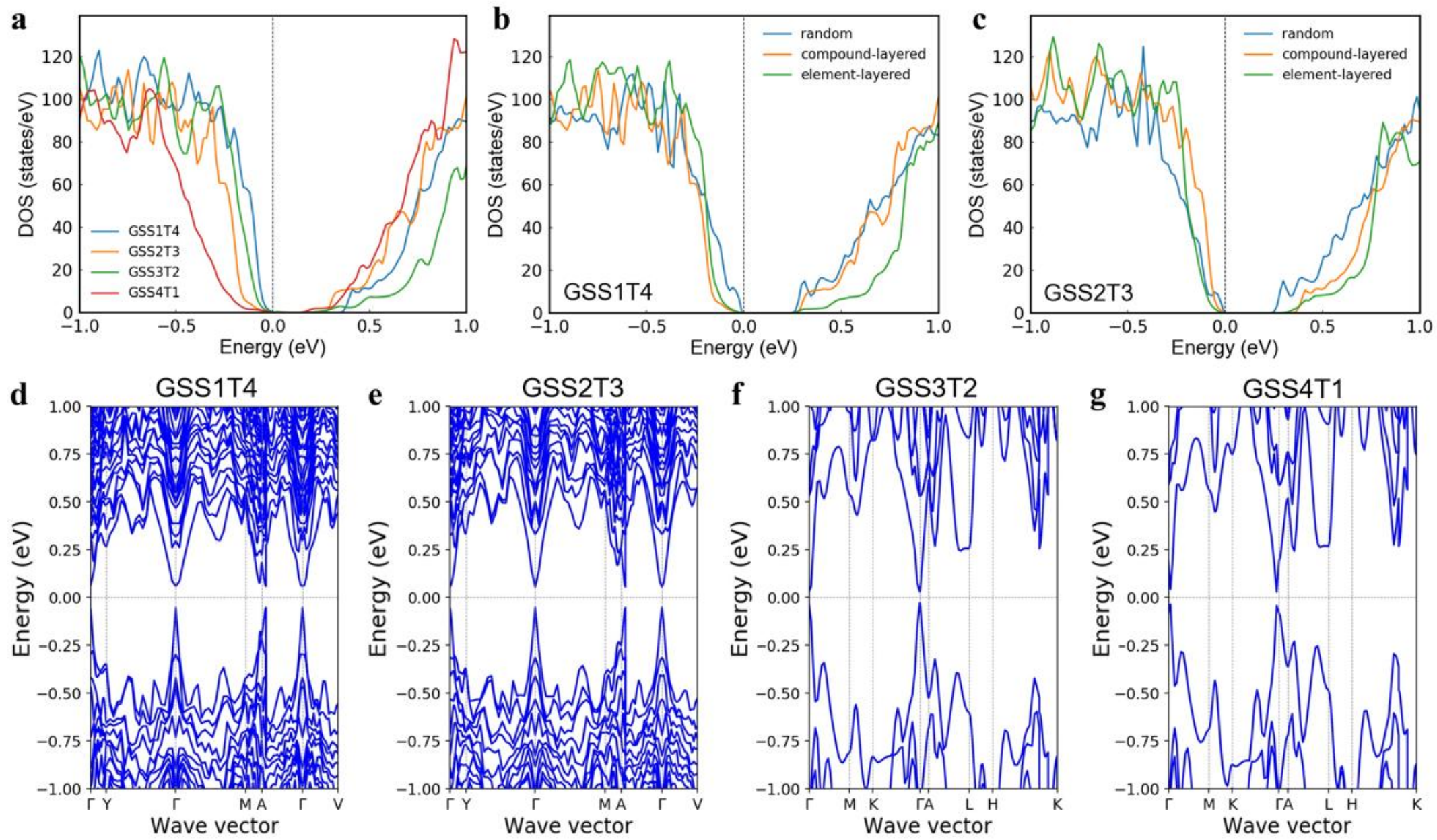

**Figure 2. a** Total density of states of $Ge_2Sb_2Se_xTe_{5-x}$ calculated based on the most stable atomic configurations. **b and c** Comparison of the density of states in random, compound-layered, and element-layered structures, for GSS1T4 and GSS2T3, respectively. **d-g** Band structure of $Ge_2Sb_2Se_xTe_{5-x}$ (GSS1T4, GSS2T3, GSS3T2 and GSS4T1) computed using the most stable atomic configurations.

Given the established vital role of antibonding states in property tuning of chalcogenide phase-change materials,[26-28] here its correlation with the distinct local ordering in GSS1T4 and GSS2T3 is probed. To quantitatively evaluate the bonding characteristics, crystal orbital Hamilton population (COHP) analysis is performed, by decomposing the electronic states into bonding, nonbonding, and antibonding contributions. Within this framework, negative COHP values represent bonding interactions, whereas positive values correspond to antibonding interactions. The energy-integrated COHP (ICOHP) is further employed to quantify the overall bonding strength. For clarity, the minus COHP and ICOHP is presented throughout this work, thus a larger value of –COHP/-ICOHP represents stronger bonding. As shown in Figure 3, for both of GSS1T4 and GSS2T3, in the element-layered structures, a pronounced peak appears at the antibonding states near the VBM, thereby reducing their overall bonding stability. And the compound-layered structures exhibit higher -ICOHP values, indicating an overall enhanced bonding and thus more significant stability. A more detailed layer-resolved COHP analysis (insets in Figure 3) further reveals that the antibonding peaks originate mainly from the Sb-Se bonds within the Ge-Se-Sb triple layers in the element-layered structures.

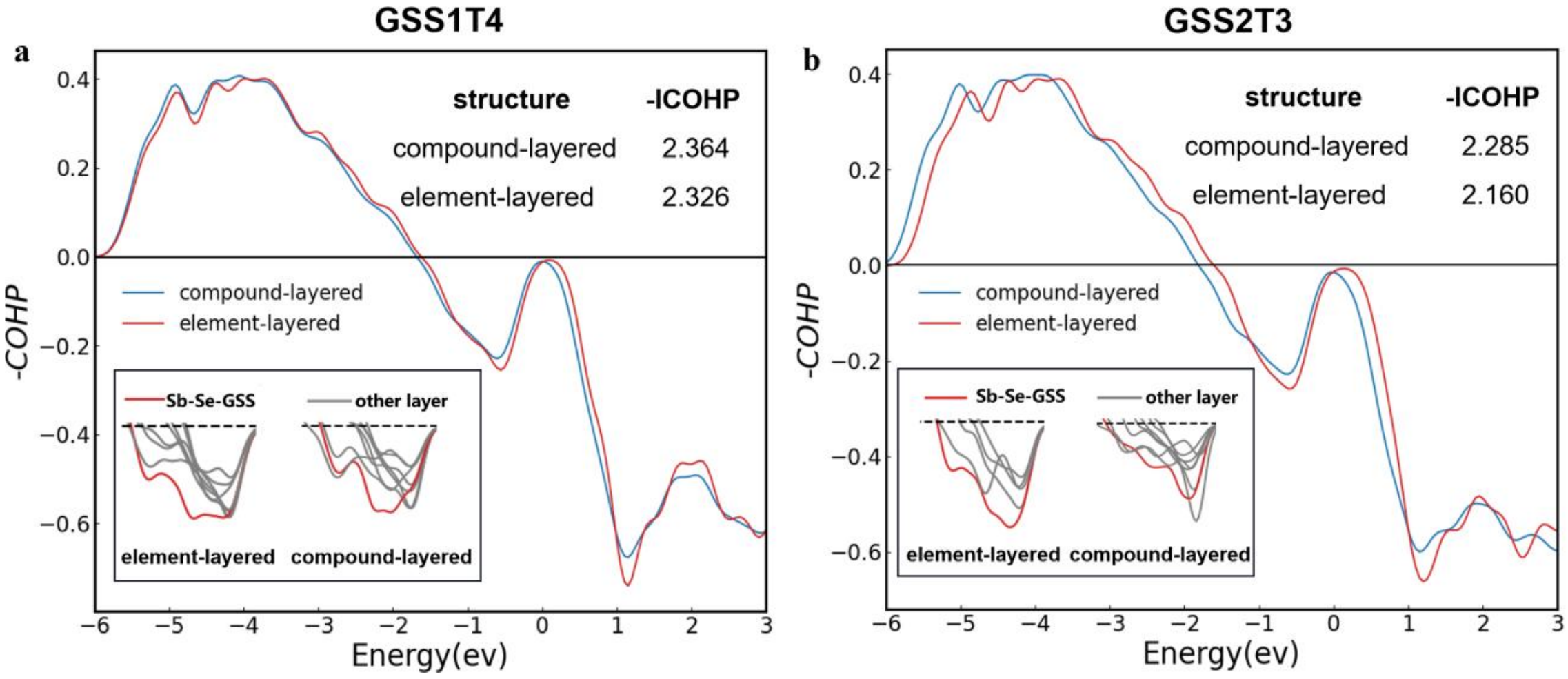

**Figure 3**. **a-b** -COHP and -ICOHP analysis of the compound-layered and element-layered structure for GSS1T4 and GSS2T3. The inset illustrates the -COHP near the Fermi level, highlighting the differences in antibonding interactions among distinct intra-layer bonds. Here, Sb-Se-GSS denotes the Sb-Se bonds within the Ge-Se(Te)-Sb triple layers.

As for the application in the field of optoelectronic fields, the optical property is another key factor for the materials design, which is studied in this section. To validate the reliability of the present computational methodology for optical property, refractive index ($n$) and extinction coefficient ($\kappa$) are calculated for $Ge_2Sb_2Se_4Te_1$ (GSS4T1), the most extensively studied composition in the GSST family. As shown in Figure S8, the calculated optical spectra of the bulk GSS4T1 are in overall agreement with the available experimental measurements using the film sample, supporting the reliability of the computational approach adopted in this work.

Then the optical response are investigated for the GSS1T4 and GSS2T3 systems, as shown in Figure 4. The results show that, as the Se concentration increases, a systematic reduction in both $n$ and $\kappa$ over the spectrum of 500-2500 nm is observed, which is consistent with previous theoretical studies on compositionally tuned optical properties[29] and the experimental observation[10]. Importantly, when comparing the compound-layered structure with the traditional element-layered structure, although the trends of $n$ and $\kappa$ with respect to wavelength are generally similar ($n$ decreases with increasing wavelength, while $\kappa$ shows a slight increase), the extinction coefficient of the compound-layered structure is consistently lower, indicating reduced optical loss. The optical response is intrinsically connected with numerous electronic features of the materials, including bandgap, DOS at the band edge, and localization of the electronic states, etc. In the present systems, for the compound-layered structure, the slight increasing of the bandgap (Figure 2b, c) can induce its decreasing extinction coefficient ($\kappa$) compared to the case of element-layered ones, although its higher DOS near the VBM or CBM tends to increase $\kappa$.

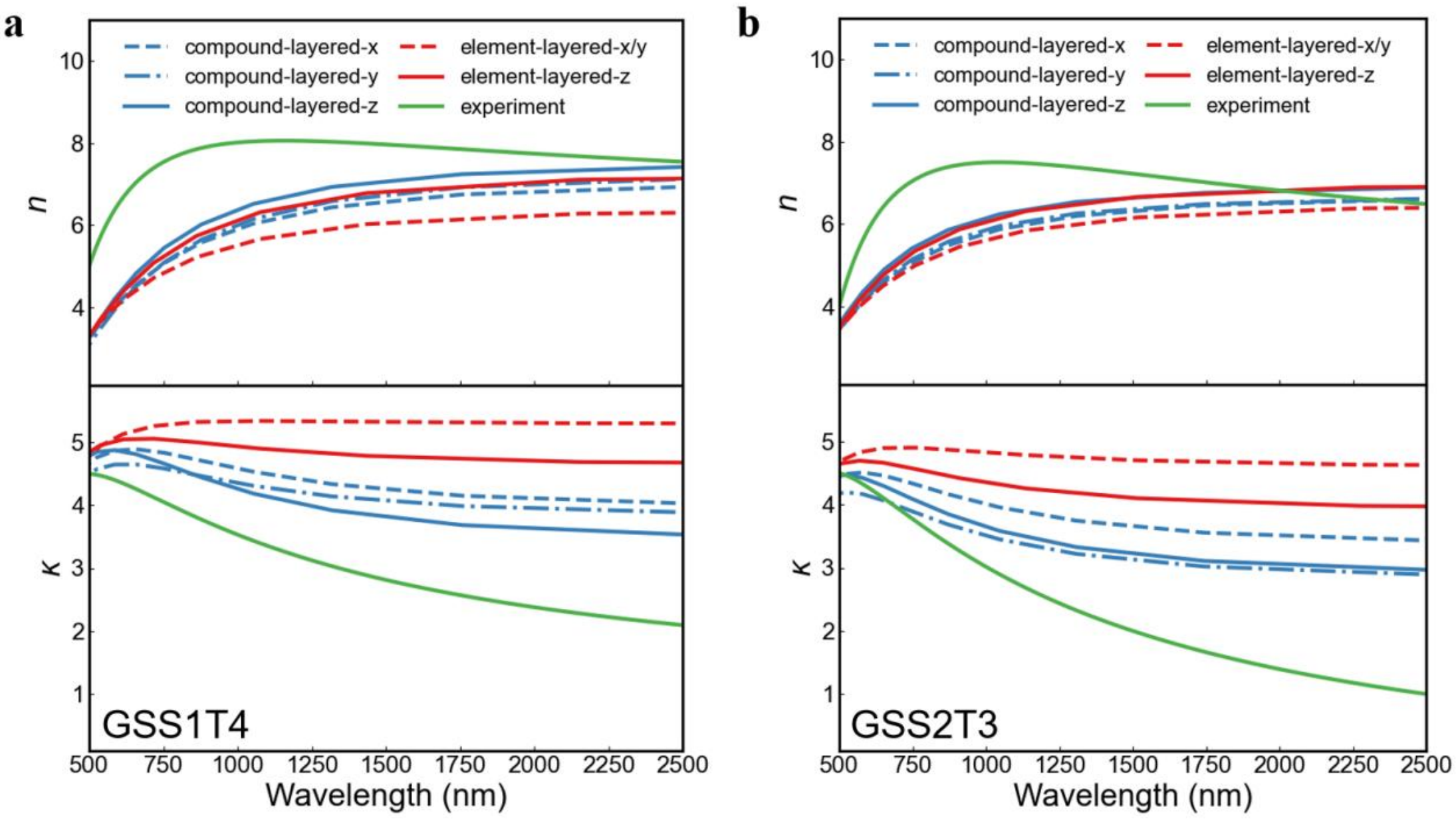

**Figure 4.** Refractive index ($n$) and extinction coefficient ($\kappa$) of **a** $Ge_2Sb_2Se_1Te_4$ and **b** $Ge_2Sb_2Se_2Te_3$. Calculated results for the anisotropic compound-layered structures are shown along three Cartesian directions (compound-layered-x, -y, and -z). For the element-layered structure, which is isotropic in the x/y plane, the in-plane values are denoted as element-layered-x/y, and the element-layered-z represents the data along z

direction. The experimental data is taken from Ref. 10.

More importantly, as shown in Figure 4, for both the refractive index $n$ and the extinction coefficient $\kappa$, our proposed compound-layered model effectively reduces the discrepancy between theoretical calculations and experimental measurements[10]. Again it should be noted that our calculations are based on bulk models, whereas the experimental data were acquired from the thin-film samples[10], which introduces inherent variations due to size effects, surface and interface contributions, residual stresses from substrate constraints, and possible structural defects,[30,31] all of which can affect the measured optical responses. Consequently, some deviation between theory and experiment remains. Nevertheless, the improved agreement with experimental data provides compelling evidence for the presence of the locally ordered compound-layered structures in actual GSST materials, rather than the idealized element-layered configurations commonly assumed in prior studies.

In conclusion, by combining DFT calculations with MC simulations, we reveal a universal tendency toward local structural ordering in $Ge_2Sb_2Se_xTe_{5-x}$ systems. Notably, $Ge_2Sb_2Se_1Te_4$ and $Ge_2Sb_2Se_2Te_3$ favor the formation of novel compound-layered configurations featuring in-layer $SeTe_2$/$Se_2Te$ stoichiometry, which differ fundamentally from the conventional element-layered structures with separated Se and Te layers. Thermodynamic analysis based on order – disorder transition temperatures confirms that these compound-layered structures remain stable up to ~370 K or higher, depending on Se content. COHP analysis further indicates that reduced antibonding interactions play a key role in stabilizing these configurations. Importantly, compared with the element-layered counterparts, the compound-layered structures exhibit a significantly reduced extinction coefficient and a slightly enhanced refractive index, achieving much better agreement with experimental observations. Overall, this work clarifies the ground-state configurations of GSST materials and highlights the crucial role of local chemical ordering in determining their optoelectronic properties, providing valuable insights for atomic-scale design of advanced optoelectronic devices. While the present simulations reveal a clear tendency toward local chemical ordering, the finite supercell size constrained by the cost of DFT calculations may limit the accurate description of possible long-range ordering phenomena. Recent advances in machine-learning interatomic potentials[32-35] offer a promising route to overcome this limitation by enabling atomistic simulations at substantially larger length scales.

**Acknowledgements**

This work is financially supported by the National Key R&D Program of China (2022ZD0117601), and National Natural Science Foundation of China (No. 52173216, 52222101, 52332005).

**Conflict of Interest Statement**

The authors declare no conflict of interest.

## References

[1] A. Sebastian, T. Tuma, N. Papandreou et al., "Temporal correlation detection using computational phase-change memory," Nat. Commun. 8, 1115 (2017).

[2] Z. Quan, Y. Wan and J. Wang, "On-chip ultra-compact nonvolatile photonic synapse," Appl. Phys. Lett. 121, 171101 (2022).

[3] W. Yuan, Y. Lu, L. Lu, R. Wang, Y. Weng, L. You, F. Zheng et al., "Electrode area dependent switching behavior of $Ge_2Sb_2Se_4Te_1$ phase change material driven by narrow voltage pulse," Appl. Phys. Lett. 122, 243503 (2023).

[4] C. Yoo, H.-K. Shin, S. S. Han, S. Lee, C. W. Lee, Y.-J. Song, T.-S. Bae, S. J. Yoo, J. Cao, J. H. Kim, H.-J. Lee, H.-S. Chung, Y. Jung et al., "Wafer-scale freestanding monocrystalline chalcogenide membranes by strain-assisted epitaxy and spalling," Nano Lett. 24, 12823–12831 (2024).

[5] J. Barnett, L. Wehmeier, A. Heßler, M. Lewin, J. Pries, M. Wuttig, J. M. Klopf, S. C. Kehr, L. M. Eng and T. Taubner, "Far-infrared near-field optical imaging and Kelvin probe force microscopy of laser-crystallized and -amorphized phase change material $Ge_3Sb_2Te_6$," Nano Lett. 21, 9012–9020 (2021).

[6] R. Audhkhasi, V. Tara, M. Klein, A. Tang, R. Chen, S. Vangala, J. R. Hendrickson and A. Majumdar, "Electrically reconfigurable nonvolatile flatband absorbers in the mid-infrared with wide spectral tuning range," Nano Lett. 25, 13533–13538 (2025).

[7] K. K. Du, Q. Li, Y. B. Lyu et al., "Control over emissivity of zero-static-power thermal emitters based on phase-changing material GST," Light Sci. Appl. 6, e16194 (2017).

[8] M. Wuttig and N. Yamada, "Phase-change materials for rewriteable data storage," Nat. Mater. 6, 824–832 (2007).

[9] M. Wuttig, H. Bhaskaran and T. Taubner, "Phase-change materials for non-volatile photonic applications," Nat. Photonics 11, 465–476 (2017).

[10] Y. Zhang, J. B. Chou, J. Li et al., "Broadband transparent optical phase change materials for high-performance nonvolatile photonics," Nat. Commun. 10, 4279 (2019).

[11] Z. Tang et al., "$Ge_2Sb_2Se_4Te$-based optical switch with ultra-high contrast ratio by multilayer Fabry–Perot cavity," Adv. Sci. 12, 2412499 (2025).

[12] X. Lian et al., "Systematic study on electronic, mechanical, and thermal transport properties of germanium antimony selenide telluride alloy by a first-principles approach," ACS Appl. Energy Mater. 6, 7919–7927 (2023).

[13] D. Sahoo and R. Naik, "GSST phase change materials and its utilization in optoelectronic devices: A review," Mater. Res. Bull. 148, 111679 (2022).

[14] A. H. A. Nohoji, P. Keshavarzi and M. Danaie, "High-performance all-optical photonic crystal synapse based on Mach–Zehnder interferometer and directional coupler utilizing GSST phase-change material," Results Phys. 69, 108338 (2025).

[15] X. Yuan, Z. Wei, Q. Ma, W. Ding and J. Guo, "Multitask learning deep neural networks enable embedded design of active metamaterials," ACS Appl. Mater. Interfaces 16, 26500–26511 (2024).

[16] N. A. Obile et al., "Time-resolved temperature mapping leveraging the strong thermo-optic effect in phase-change materials," ACS Photonics 10, 3576–3585 (2023).

[17] Z. Fang, R. Chen, V. Tara and A. Majumdar, "Non-volatile phase-change materials for programmable photonics," APL Mater. 11, 0165309 (2023).

[18] K. Aryana, Y. Zhang, J. A. Tomko, M. S. B. Hoque, E. R. Hoglund, D. H. Olson, J. C. Read, C. Ríos, J. Hu and P. E. Hopkins, "Suppressed electronic contribution in thermal conductivity of $Ge_2Sb_2Se_4Te$," Nat. Commun. 12, 7187 (2021).

[19] A. Lotnyk et al., "Real-space imaging of atomic arrangement and vacancy layers ordering in laser crystallised $Ge_2Sb_2Te_5$ phase change thin films," Acta Mater. 105, 1–8 (2016).

[20] L. M. Vogl, S. Chen, P. Schweizer, X. Jin, S. Q. Yu, J. Liu and A. M. Minor, "Identification of short-range ordering motifs in semiconductors," Science 389, 1342–1346 (2025).

[21] Z. Ma, Y. Luo, J. Dong, Y. Liu, D. Zhang, W. Li and M. G. Kanatzidis, "Synergistic performance of thermoelectric and mechanical in nanotwinned high-entropy semiconductors $AgMnGePbSbTe_5$," Adv. Mater. 36, 2407982 (2024).

[22] T. Zhang, L. Zhu, H. Liu, J. Zhou and Z. Sun, "Ordering phenomena in ternary transition-metal dichalcogenides: Critical role of lattice symmetry and vdW interaction," Mater. Genome Eng. Adv. 1, e7 (2023).

[23] H. Liu, L. Zhu, J. Zhou et al., "Competing sublattice short-range orders and gap state engineering in multicomponent transition-metal dichalcogenide," npj Comput. Mater. 11, 189 (2025).

[24] C. Toher, C. Oses, D. Hicks and S. Curtarolo, "Unavoidable disorder and entropy in multi-component systems," npj Comput. Mater. 5, 69 (2019).

[25] A. Jain, S. P. Ong, G. Hautier, W. Chen, W. D. Richards, S. Dacek, S. Cholia, D. Gunter, D. Skinner, G. Ceder and K. A. Persson, "The Materials Project: A

materials genome approach to accelerating materials innovation," APL Mater. 1, 011002 (2013).

[26] Y. Huang, J. Zhou, L. Peng, K. Li, S. R. Elliott and Z. Sun, "Antibonding-induced anomalous temperature dependence of the band gap in crystalline $Ge_2Sb_2Te_5$," J. Phys. Chem. C 125, 19537–19543 (2021).

[27] Y. Zhou, L. Sun, G. M. Zewdie, R. Mazzarello, V. L. Deringer, E. Ma and W. Zhang, "Bonding similarities and differences between Y–Sb–Te and Sc–Sb–Te phase-change memory materials," J. Mater. Chem. C 8, 3646–3654 (2020).

[28] S. Hu, J. Xiao, J. Zhou, S. R. Elliott and Z. Sun, "Synergy effect of co-doping Sc and Y in $Sb_2Te_3$ for phase-change memory," J. Mater. Chem. C 8, 6672–6679 (2020).

[29] H. Zhang, X. Wang and W. Zhang, "First-principles investigation of amorphous Ge–Sb–Se–Te optical phase-change materials," Opt. Mater. Express 12, 2497–2508 (2022).

[30] S. Ali, R. Magnusson, O. V. Pshyk, A. Le Febvrier et al., "Effect of O/N content on the phase, morphology, and optical properties of titanium oxynitride thin films," J. Mater. Sci. 58, 10975–10985 (2023).

[31] D. E. Aspnes, "Optical properties of thin films," Thin Solid Films 89, 249–262 (1982).

[32] L. Zhang, J. Han, H. Wang, R. Car and W. E, "DeePMD-kit: A deep learning package for many-body potential energy representation and molecular dynamics," Comput. Phys. Commun. 228, 178–184 (2018).

[33] E. V. Podryabinkin, K. Garifullin, A. V. Shapeev and I. S. Novikov, "MLIP-3: Active learning on atomic environments with moment tensor potentials," J. Chem. Phys. 159, 084112 (2023).

[34] B. Deng, P. Zhong, K. Jun, J. Riebesell, K. Han, C. J. Bartel and G. Ceder, "CHGNet as a pretrained universal neural network potential for charge-informed atomistic modelling," Nat. Mach. Intell. 5, 1031–1041 (2023).

[35] Y. Zhou, W. Zhang, E. Ma and V. L. Deringer, "Device-scale atomistic modelling of phase-change memory materials," Nat. Electron. 6, 746 – 754 (2023).